\documentclass[a4paper]{article}

\usepackage{INTERSPEECH2019}

\usepackage{graphicx}
\usepackage{color, soul, multirow}
\usepackage{subfig}
\usepackage{bm}
\usepackage{verbatim}
\usepackage{amssymb}
\usepackage{pifont}
\usepackage{float}
\usepackage{amsmath}
\usepackage{booktabs}
\usepackage{pifont}
\usepackage{tikz}
\usepackage{pgfplots}
\newcommand{\cmark}{\ding{51}}%
\newcommand{\xmark}{\ding{55}}%

\title{Cross-Attention End-to-End ASR for Two-Party Conversations}
\name{Suyoun Kim$^1$, Siddharth Dalmia$^2$, Florian Metze$^2$}
\address{
  $^1$Electrical \& Computer Engineering\\
  $^2$Language Technologies Institute, School of Computer Science\\
  Carnegie Mellon University \\
  Pittsburgh, PA 15213; U.S.A.
}
\email{\{suyoung1{\textbar}sdalmia{\textbar}fmetze\}@andrew.cmu.edu}

\begin{document}

\maketitle

\begin{abstract}
  We present an end-to-end speech recognition model that learns interaction between two speakers based on the turn-changing information. Unlike conventional speech recognition models, our model exploits two speakers’ history of conversational-context information that spans across multiple turns within an end-to-end framework. Specifically, we propose a speaker-specific cross-attention mechanism that can look at the output of the other speaker side as well as the one of the current speaker for better at recognizing long conversations. We evaluated the models on the Switchboard conversational speech corpus and show that our model outperforms standard end-to-end speech recognition models.
\end{abstract}

\section{Introduction}
    Contextual information plays an important role in automatic speech recognition (ASR), especially in processing a long conversation since semantically related words, or phrases often reoccur across sentences. Typically, a long contextual information is only modeled in the language model (LM) which is trained only on text data separately \cite{mikolov2010recurrent, mikolov2012context, wang2015larger, ji2015document, liu2017dialog, xiong2018session}, then the language model combined with the acoustic model trained on isolated utterances in decoding phase. Such disjoint modeling process may not exploit the useful contextual information fully. 
    
    Several recent work attempted to use the contextual information with a recent progress \textit{end-to-end} speech recognition framework, promises to integrate all available information into a single model, which is jointly optimized \cite{kim2018dialog, kimsituation, kim2019acoustic, kim2019acl, pundak2018deep, alon2018contextual}. In \cite{kim2018dialog, kimsituation, kim2019acoustic, kim2019acl}, they proposed to use conversational-context embeddings which encodes previous utterance prediction to predict the current output token and showed promising results, however they did not consider the two-speaker interaction in a conversation. In \cite{pundak2018deep, alon2018contextual}, they proposed to use phrase list (e.g. \textit{song list, contact list} with attention mechanism and showed significant performance improvement, however, the model assumes that it has access to the phrase list at inference. 

    In this work, we create a cross-attention end-to-end speech recognizer capable of incorporating the two speaker conversational-context information to better process long conversations. Specifically, we first propose to use an additional attention mechanism to find more informative utterance representation among the multiple histories and focus more on it. Additionally, we propose to use LSTM with attention mechanism that specifically track the interactions between two speakers. We evaluate our model on the Switchboard conversational speech corpus \cite{swbd, godfrey1992switchboard}, and show that our model outperforms the standard end-to-end speech recognition model.

\begin{figure*}[t]
    \centering
    \centerline{\includegraphics[height=3.2in]{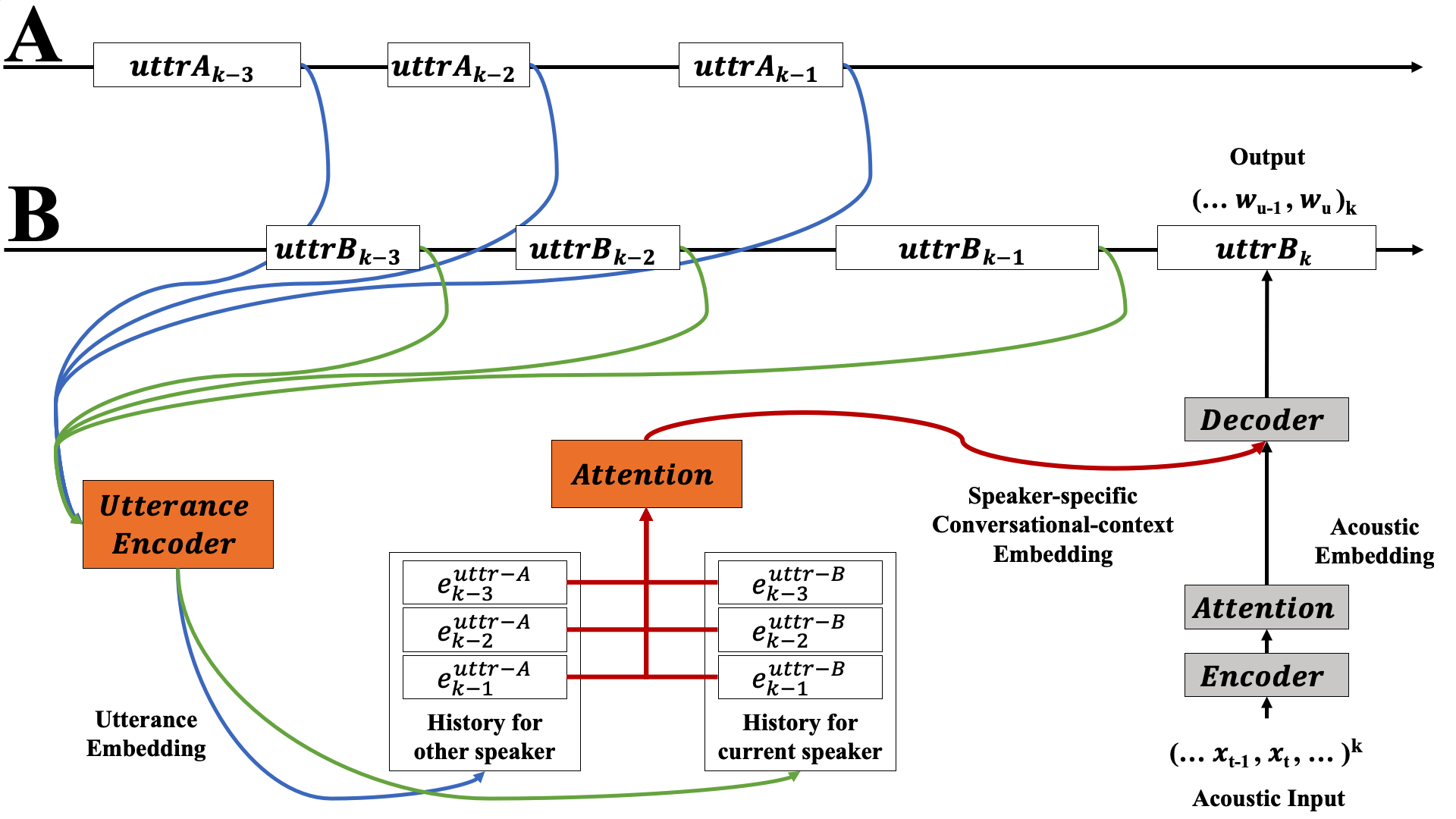}}
    \caption{Overall architecture of our proposed end-to-end speech recognition model using speaker-specific conversational-context embedding generated from the utterance history of both other speaker and current speaker. Our framework uses the utterance encoder to generate the utterance embedding, and uses additional attention mechanism for the multiple utterance embeddings from two speakers. Then, speaker-specific conversational-context embedding is forwarded to the conversational end-to-end speech recognizer. All of these model is a single network and trained in end-to-end manner.}
    \label{fig:arch}
\end{figure*}

\section{Related work}
    Several recent studies have considered to incorporate the context information within a end-to-end speech recognizer \cite{pundak2018deep,alon2018contextual}. In contrast with our method which uses a conversational-context information for processing a long two-speaker conversation, their methods use distinct phrases (i.e. play a song) with an attention mechanism in specific tasks, \textit{contact names, songs names, etc}. Their model assumes there exists such a list of phrases at inference.  
    
    In study \cite{liu2017dialog}, they proposed contextual RNN language models that track the interactions between speakers by using additional RNN. In contrast with our work, they built a language model which is trained only on text corpus.  
    
    Several recent studies have considered to embed a longer context information within a end-to-end framework \cite{kim2018dialog, kim2019acoustic, kim2019acl}. In contrast with our method, we consider multiple utterance histories for each speaker and use the attention mechanism with LSTM for learning the interaction between the two speakers. 
    
    \vspace{0.5cm}
    
\section{Model}
    In this section, we first review conversational-context aware end-to-end speech recognition model \cite{kim2018dialog, kim2019acoustic, kim2019acl}. We then present our proposed cross-attention end-to-end speech recognition model for processing two speaker conversations.  

\subsection{Encoder for history of the utterances}
\label{sec:utterance_encoder}
    
    The key idea of the conversational-context aware end-to-end speech recognition model \cite{kim2018dialog, kim2019acoustic, kim2019acl} is to use the history of utterance representation for predicting each output token. In order to obtain the history of utterance representation, an additional utterance encoder is used within the decoder network in the standard sequence-to-sequence framework \cite{bahdanau2014neural, chorowski2014end, chorowski2015attention, chan2015listen, kim2017joint}. This utterance encoder maps the variable-length output tokens from multiple preceding utterances into the fixed-length single vector, then it is fed into the decoder at each output time step. 

    Let we have $K$-utterances and acoustic features $(x_1, \cdots, x_T)^k$ and output word sequence, $(w_1, \cdots, w_U)$ for each $k$-th utterance. At each output step $u$ in prediction of $k$-th utterance, decoder predicts the word distribution by conditioning on 1) acoustic embedding  ($e_{a}^{k}$) from \texttt{Encoder}, 2) previous word embedding ($e^{u-1}_{w}$), and additionally 3) conversational-context embedding ($e^{k}_{c}$):

    \begin{align}
        \label{eq:loss}
        e^{k}_{a}       = & \texttt{Encoder}(\mathbf{x^k}) \\
        \mathbf{w^k_u} \sim & \texttt{Decoder}(e^{k}_{c}, e^k_{w}, e^{k}_{a})
    \end{align}    

\vspace{1cm}
    
    For representing the utterance into the fixed length vector representation, $e^{k}_{c}$, we can use 1) mean of one-hot word vectors, $\texttt{mean}(e^{k-1}_{w} + \cdots + e^{k-1}_{w})$, or 2) mean of external word embeddings, (i.e. Word2Vec \cite{mikolov2012context}, GloVe \cite{pennington2014glove}, fastText \cite{bojanowski2017enriching}, etc), or 3) external sentence embedding resource (i.e. ELMo \cite{Peters:2018}, BERT \cite{devlin2018bert}, etc). In this work, we are focusing on learning interactions of two speaker conversation rather than focusing on exploring the method of conversational-context representation, we thus use 3) BERT representation method in all of our experiments. 
    
    In order to pass the conversational-context across mini-batches during training and decoding, the dataset is serialized based on their onset times and their dialogs rather than random shuffling of data. Then, we create the mini-batches that contain only one utterance of each dialog to pass the context embedding to the next mini-batches properly.

%\vspace{0.5cm}

\subsection{Cross-Attention for two speakers' conversation}
    Although the previously proposed conversational-context aware end-to-end ASR model exploits the utterance of history as an additional information, there are two limitations. First, the model does not consider multiple speaker case and interaction between them, which is common and useful information in processing the conversation. Second, the model simply concatenates the multiple utterance embeddings then projects to a fixed dimensional vector or use the mean of the multiple utterance embeddings, so it does not explicitly allow to attend more on the more important utterance embedding. 
    
    Based on the two observations above, we therefore propose two methods to extend the current conversational-context aware end-to-end ASR for processing two-party conversations. The overall architecture of our proposed model is described in Figure \ref{fig:arch}. Specifically, our model works as follows. 
    
    We first represent an utterance to a fixed-length vector representation as described in Section \ref{sec:utterance_encoder}. The utterance encoder maps the sequence of one-hot word vectors to the single, dense vector, the utterance embedding. 
    
    Next, we create a queue for each speaker to store the history of utterance embeddings as described in Figure \ref{fig:arch}. In this work, we consider two-speaker conversations and assume that the turn-changing information is known so that the utterance embeddings can be stored separately. 
    
    We then use an attention mechanism to generate speaker-specific conversational-context embedding given the history of \textit{what other speaker said} and the history of \textit{what current speaker said}. Note that, based on what current speaker is, we swap the queues properly. We propose two methods to generate the attended context embeddings.

    \subsubsection{Attention over each speaker's utterance history}
    
    First method is simply using an additional attention mechanism over the utterance embeddings. Given the $N$- size of the utterance history for speaker A, $e_{k-N}^{u-A}, \cdots , e_{k-1}^{u-A}$, the conversational-embedding, $\texttt{att\_e}_{k}^{A}$, is generated as follows: 
    \begin{align}
        \mathbf{G}_{k}^{A} & = \texttt{tanh} (\mathbf{W} e_{k-N:k-1}^A + \mathbf{b}) \\
        \alpha_{k}^{A} & = \texttt{softmax} (\mathbf{w}^T \mathbf{G}_{k}^{A} + \mathbf{b}) \\
        \texttt{att\_e}_{k}^{A} & = \sum_{N} a_{k}^{A} \odot e^A_{k-N:k-1}
    \end{align}
    where $\mathbf{W}$, $\mathbf{b}$ are trainable parameters. $e_{k}^{B}$ is generated in the same way. 
    
%\vspace{0.5cm}
    \subsubsection{Cross-attention between two speakers' utterance history}
    
    Second method is using \texttt{LSTM} with an attention mechanism. Inspired by the \texttt{matchLSTM} model which has been widely used in question answering tasks and natural languge inferance (NLI) \cite{wang2015learning, wang2016machine}, we consider to track the interaction between two speakers sequentially, by attending \textit{what other speaker said} at each utterance timestamp. The idea of the \texttt{matchLSTM} is to attempt to take the question (premise) and the passage (hypothesis) along with an answer pointer \cite{vinyals2015pointer} pointing to the start and the end of the answer to make predictions. The \texttt{matchLSTM} tries to obtain a question-aware representation of the passage, by attending over the representations of the question tokens for each token in passage. 
    
    The key difference in our work is that the question (premise) is a sequence of utterance-embedding from other speaker (\textit{what other speaker said}), and passage (hypothesis) is a sequence of utterance-embedding from current speaker (\textit{what current speaker said}). The embedding of \textit{what current speaker said} takes into consideration the alignment between the \textit{what current speaker said} and \textit{what other speaker said}. 
    
    By using \texttt{matchLSTM} over the first simple attention method, there are two benefits -
    %\vspace{0.5cm}
    \begin{itemize}
        \item First, the model is able to handle a longer utterance-history 
        \item Second, the model can learn the interaction between the two speakers, as the \texttt{matchLSTM} can potentially track the flow of the conversations.
    \end{itemize}

%\vspace{0.5cm}

    Specifically, the attended conversational embedding at $i$-th utterance-history step is generated as follows:
    \vspace{0.5cm}
    \begin{align*}
    \mathbf{G}_{k_i}^{A} &= \texttt{tanh}(\mathbf{W} e_{k-N:k-1}^B + \mathbf{V} e_{i}^A + \mathbf{U} \mathbf{h}_{i-1}^{A} + \mathbf{b}) \\
    \alpha_{k_i}^{A} &= \texttt{softmax}(\mathbf{w}^T \mathbf{G}_{k_i}^{A} + b)
    \end{align*}
    \vspace{0.5cm}
    where $\mathbf{W}, \mathbf{V}, \mathbf{U}, \in \mathbb{R}^{h \times h}$, $\mathbf{b} \in \mathbb{R}^h$, $b \in \mathbb{R}$ are trainable parameters. Each hidden state $\mathbf{h}_{i-1} \in \mathbb{R}^h$ comes from the output of the \texttt{matchLSTM} that is fed the following $\mathbf{z}_i$ as input.
    \vspace{0.5cm}
    \begin{align*}
        \mathbf{z}_i^{A}    &= [e_{i}^A, e_{k-N:k-1}^B \odot \alpha_{k_i}^{A}] \\
        \mathbf{h}_i^{A}   &= \texttt{matchLSTM}(\mathbf{z}_i^{A}, \mathbf{h}_{i-1}^{A})
    \end{align*}
    \vspace{0.5cm}
    Using a LSTM, there are $N$ such $h$-dimensional hidden states, and we take the final hidden states for our attended conversational-context embedding:
    \begin{align*}
        \texttt{att\_e}_{k}^{A} & = \mathbf{h}_{k-1}^{A}
    \end{align*}
    %
    
%\vspace{0.5cm}
    Finally, the decoder network takes the above attended conversational-context embedding, $\texttt{att\_e}_{k}^{A}$, generated from the either method, in addition to the usual inputs, acoustic embedding from encoder network and the previous word embedding in current utterance. In this work, we used the same, 100 dimension for the conversational embedding for all of our experiments.

%\vspace{0.5cm}
    
\section{Experiments}
\label{sec:exp}
\subsection{Datasets}
    We trained our model on 300 hours of two-party conversational speech corpus, the Switchboard LDC corpus (97S62), and tested on the HUB5 Eval 2000 LDC corpora (LDC2002S09, LDC2002T43). In Section \ref{sec:result}, we show separate results for the CallHome English (CH) and Switchboard (SWB) sets. We denote the number of utterances, the number of dialogs, the average number of utterances per dialog, and the number of speakers for each training, validation, evaluation sets in Table \ref{tab:data}.

\begin{table}[H]
    \caption{ Experimental dataset description. We used 300 hours of Switchboard conversational corpus. Note that any pronunciation lexicon or Fisher transcription was not used.}
    \label{tab:data}
    \begin{center}
        \resizebox{\columnwidth}{!}{
        \begin{tabular}{r|r|r|r|r}
            \toprule
            numbers                &  training  &  validation & SWB   & CH\\
            \midrule
            utterances            & 192,656    & 4,000      & 1,831     & 2,627 \\
            dialogs               & 2402       & 34         & 20        & 20    \\ 
            utterances / dialog & 80         & 118        & 92        & 131   \\ 
            speakers              & 4804       & 68         & 40        & 40    \\ 
            \bottomrule
        \end{tabular}
        }
    \end{center}
\end{table}

    The input feature for each frame which was sampled audio data at 16kHz is represented by a 83-dimensional feature vector, consisting of 80-dimensional log-mel filterbank coefficients and 3-dimensional pitch features. We used same output units in previous work \cite{kim2019acl}, consisting of 10,038 the word units and the single character units.

\begin{table*}[t!]
    \centering
    \caption{Comparison of word error rates (WER) on Switchboard 300h with standard end-to-end speech recognition models and our proposed end-to-end speech recognition models with two-speaker conversational context. Note that our baselines did not use the external language model, Fisher text data, CD phones information, or the layer-wise pre-training technique \cite{zeyer2018improved}. (The * mark denotes our estimate for the number of parameters used in the previous work.}
    \label{tab:result}
    \resizebox{\textwidth}{!}{
    \begin{tabular}{r|r|r|r|r|r|r}
        \toprule
                    &         & Number of  &   Number of &   External  & SWB & CH \\
        Model   & Output Units & trainable parameters & utterance-history & LM & (WER\%) & (WER\%) \\
        \midrule
        \textbf{Prior Models} &&& & & & \\
        LF-MMI \cite{povey2016purely} & CD phones & N/A && \cmark & 9.6 & 19.3 \\
        CTC \cite{zweig2017advances} & Char & 53M &&  \cmark & 19.8 & 32.1 \\
        CTC \cite{sanabria2018hierarchical} & Char, BPE-{300,1k,10k} & 26M && \cmark & 12.5 &23.7 \\
        CTC \cite{audhkhasi2017building} & Word (Phone init.) & N/A && \cmark & 14.6  & 23.6 \\
        Seq2Seq \cite{zeyer2018improved} & BPE-10k & 150M* && \xmark & 13.5 & 27.1 \\
        Seq2Seq \cite{palaskar2018acoustic} & Word-10k & N/A && \xmark & 23.0 & 37.2 \\
        %Seq2Seq \cite{zeyer2018improved} & BPE-1k & 150M* && \cmark & 11.8 & 25.7 \\
        \midrule
        \textbf{Our Baselines} &  &  &  &  & &  \\
        \texttt{Baseline} & Word-10k & 32M & \xmark & \xmark & 17.9	& 30.6 \\
        \textbf{Our Models}& & && & & \\
        \texttt{Attention} & Word-10k &  34M & 6    & \xmark & 16.7	& 30.0  \\
        \texttt{matchLSTM} & Word-10k &  34M & 6    & \xmark & 16.4	& 29.9  \\
         %\midrule
        \texttt{Attention} & Word-10k &  34M & 10   & \xmark & 16.6	& 30.0  \\
        \texttt{matchLSTM} & Word-10k &  34M & 10   & \xmark & 16.4	& 29.9  \\
         %\midrule
        \texttt{Attention} & Word-10k &   34M & 20  & \xmark & 16.6	& 30.0  \\
        \texttt{matchLSTM} & Word-10k &   34M & 20  & \xmark & 16.4	& 29.8  \\
        \bottomrule
    \end{tabular}
    }
\end{table*}

%\vspace{0.5cm}

\subsection{Training and decoding}
\label{sec:training}
    
    For the standard-end-to-end speech recognition model, we used joint CTC/Attention framework \cite{kim2017joint, watanabe2017hybrid} which is based on the sequence-to-sequence framework \cite{bahdanau2014neural, chorowski2014end, chorowski2015attention, chan2015listen} with using CTC \cite{graves2006connectionist} as an auxiliary objective function. We followed the same network architecture as the prior study in \cite{zhang2017very, hori2017advances}. We used CNN-BLSTM encoder and LSTM decoder. The CNN-BLSTM encoder consists of 6-layer CNN and 6-layer BLSTM with 320 cells. The LSTM decoder was 2-layer LSTM with 300 cells. Our proposed model requires approximately 2M trainable parameters for the speaker-specific attention mechanisms compared to our baseline. In Table \ref{tab:result} shows the total number of trainable parameters. 
    
    For optimization, we used AdaDelta \cite{zeiler2012adadelta} with gradient clipping \cite{pascanu2013difficulty}. We bootstrapped the training our proposed two-speaker conversational end-to-end models from the vanilla conversational models and the baseline end-to-end models. 
    
    For decoding, we used left-right beam search algorithm \cite{sutskever2014sequence} with the beam size 10. We adjusted the score by adding a length penalty since the model has a small bias for shorter utterances. The final score is normalized with a length penalty $0.1$.

    The models are implemented using the PyTorch deep learning library \cite{paszke2017automatic}, and ESPnet toolkit \cite{kim2017joint, watanabe2017hybrid, watanabe2018espnet}.

\section{Results}
\label{sec:result}
    The Table \ref{tab:result} summarizes the results of our baseline and proposed model, \texttt{Attention} and \texttt{matchLSTM}, with various number of utterance history. The \texttt{Baseline} is our baseline which is trained on isolated utterances without using conversational-context information. Our proposed models were bootstrapped the training from the vanilla conversational model, and the vanilla model was also bootstrapped the training from the \texttt{Baseline}. We found that this pre-training procedure is necessary since we need to learn the additional parameters in decoder network. 
    
    We also shows the results of prior models from other literature. Note that our baselines used smaller number of model parameters, did not use the external language model, Fisher text data, CD phones information, or the layer-wise pre-training technique \cite{zeyer2018improved}. 
    
\begin{figure}[!ht]
  \centering
  \centerline{\includegraphics[width=8cm]{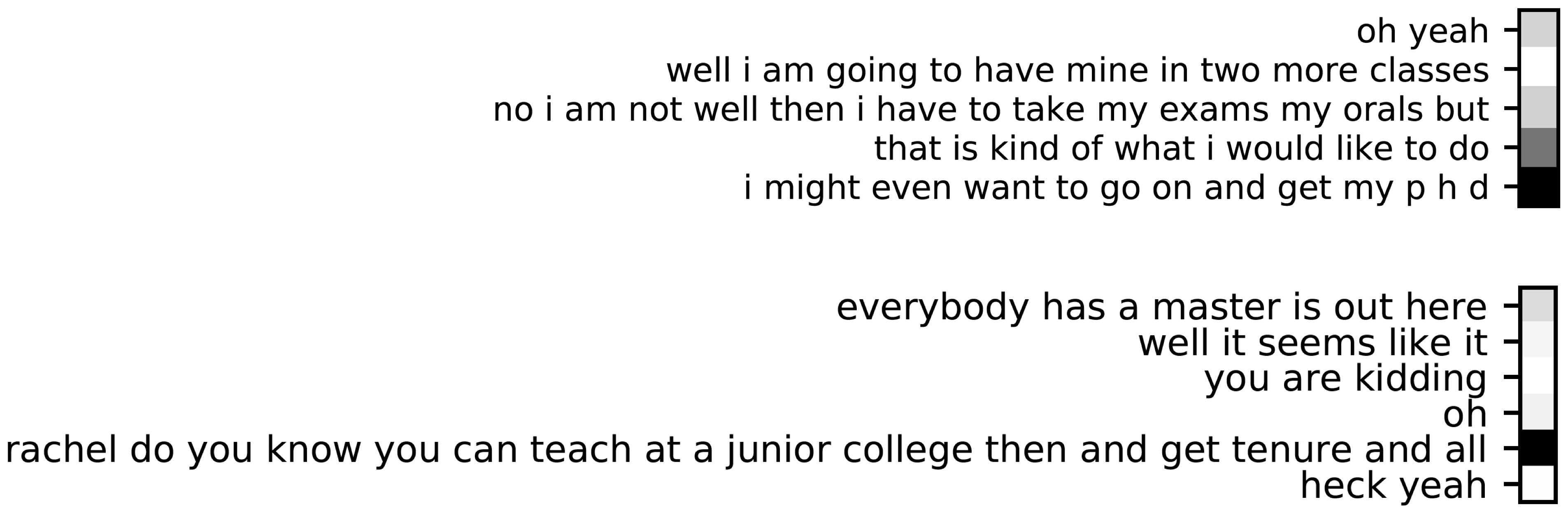}}
\caption{The attention weights over utterance-history of the speaker A (top) and the speaker B (bottom) when the model predicts the utterance (come out here to California) in evaluation set. The dark color represents higher attention weight.}
\label{fig:attention}
\end{figure}

    We first observed that our conversational models with both \texttt{matchLSTM} method and \texttt{Attention} method achieved better performance than the \texttt{baseline}. Specifically, the proposed model with \texttt{matchLSTM} using 20 utterance-history performed best, showing 16.4\% WER and 29.8\% WER in Switchboard (SWB) and CallHome (CH) evaluation sets, respectively. Figure \ref{fig:attention} visualizes the attention weights over the utterance-history. It shows that the model focuses on the long, informative utterances, rather than the short, meaningless utterances, i.e. \textit{oh}, \textit{heck yeah}, etc. 

    One noticeable thing is that the \texttt{matchLSTM} method slightly outperformed the \texttt{Attention} method in various number of utterance-history. It is possible that the \texttt{matchLSTM} method can track the flow of the conversation because it generates the better speaker-specific conversational context representation by conditioning on what others and current speaker said at each step. This point needs to be verified with additional experiments in future work.

\section{Conclusions}

    We have introduced an end-to-end speech recognizer with speaker-specific cross-attention mechanism for two-party conversations. Unlike conventional speech recognition models, our model generates output tokens conditioning on two speakers' conversational history, and consequently improves recognition accuracy of a long conversation. Our proposed speaker-specific cross-attention mechanism can look at what other speaker said in addition to what current speaker said. We evaluated the models on the Switchboard conversational speech corpus and show that our proposed model using cross-attention achieves WER improvements over the baseline end-to-end model for Switchboard and CallHome. 
    
    A future direction would be to explore the variant of the current cross-attention model, such as taking the word-level history rather than the utterance-level history. We also plan to analyze the effect of the attention mechanism to get better understanding.

\section{Acknowledgments}
    We gratefully acknowledge the support of NVIDIA Corporation with the donation of the Titan Xp GPU used for this research. This work also used the Bridges system, which is supported by NSF award number ACI-1445606, at the Pittsburgh Supercomputing Center (PSC).

\bibliographystyle{IEEEtran}
\bibliography{mybib}

\end{document}